\documentclass[onecolumn,superscriptaddress,
showpacs,pra]{revtex4}
\usepackage{mathrsfs}
\usepackage{amssymb, amsbsy, amsmath, latexsym, dsfont, array, layout, graphicx,mathrsfs,color}

\newcommand{\ket}[1]{\left|{#1}\right\rangle}
\newcommand{\bra}[1]{\left\langle{#1}\right|}

\begin{document}

\title{A one-dimensional quantum walk with multiple-rotation on the coin}
\author{Peng Xue
}\email{gnep.eux@gmail.com} \affiliation{Department of Physics,
Southeast University, Nanjing 211189, China} \affiliation{State Key
Laboratory of Precision Spectroscopy, East China Normal University,
Shanghai 200062, China}
\author{Rong Zhang}
\affiliation{Department of Physics, Southeast University, Nanjing
211189, China}
\author{Hao Qin}
\affiliation{Department of Physics, Southeast University, Nanjing
211189, China}
\author{Xiang Zhan}
\affiliation{Department of Physics, Southeast University, Nanjing
211189, China}
\author{Zhihao Bian}
\affiliation{Department of Physics, Southeast University, Nanjing
211189, China}
\author{Jian Li}
\affiliation{Department of Physics, Southeast University, Nanjing
211189, China}
\begin{abstract}
We introduce and analyze a one-dimensional quantum walk with two
time-independent rotations on the coin. We study the influence on
the property of quantum walk due to the second rotation on the coin.
Based on the asymptotic solution in the long time limit, a ballistic
behaviour of this walk is observed. This quantum walk retains the
quadratic growth of the variance if the combined operator of the
coin rotations is unitary. That confirms no localization exhibits in
this walk. This result can be extended to the walk with multiple
time-independent rotations on the coin.
\end{abstract}


\maketitle

Quantum walks (QWs) are valuable in diverse areas of science, such
as quantum algorithms~\cite{K03,A03,CC+03,SKW03,BW10,FMB11}, quantum
computing~\cite{C09,CGW13,LC+10}, transport in biological
systems~\cite{L11,HSW10} and quantum simulations of physical system
and important phenomena such as Anderson
localization~\cite{A58,W12,K10,SK10,ZXT14,SC+11,CO+13,XQT14}, Bloch
oscillation~\cite{W04,BN+06,GA+13,XQTB14} and non-trivial
topological structure~\cite{KR+10,A13,KB+12}.

%In discrete-time QWs, the coin degrees of freedom offer the
%possibility for a wider range of controls over the evolution of the
%walk than are available in the continuous-time QWs.

We study one possible route to the localization effect for the QW on
the line: the use of multiple-rotation on the coin in order to
change interference pattern between paths~\cite{PG+14}. We find
exact analytical expressions for the time-dependence of the first
two moments $\langle x\rangle_t$ and $\langle x^2\rangle_t$, show
the behaviour of QWs with two time-independent rotations on the coin
and present that a ballistic behaviour instead of localization is
observed. This result can be extended to the walk with multiple
time-independent rotations on the coin.

The unitary operator for single-step of this QW with two
time-independent rotations on the coin is
\begin{equation}
U(\theta,\phi)=T R_{{\bf n}_1}(\phi)R_{{\bf n}_2}(\theta).
\end{equation}
The two rotations on the coin are
\begin{align}
R_{{\bf n}_{1(2)}}(\theta)=e^{i\theta{\bf n}_{1(2)}\cdot{\bm
\sigma}} =\left(
            \begin{array}{cc}
              \cos\theta+in_{z1(2)}\sin\theta & (in_{x1(2)}+n_{y1(2)})\sin\theta \\
              (in_{x1(2)}-n_{y1(2)})\sin\theta & \cos\theta-in_{z1(2)}\sin\theta \\
            \end{array}
          \right),
          \label{eq:rotation}
\end{align}
where $\bm{\sigma}=(\sigma_x,\sigma_y,\sigma_z)^\text{T}$ is the
vector of Pauli matrices. The rotations are followed by a
conditional position shift operator
\begin{equation}
T=S\otimes\mathcal{P}_0+S^\dagger\otimes\mathcal{P}_1,
\end{equation}
where $\mathcal{P}_0=\ket{0}\bra{0}$ and
$\mathcal{P}_1=\ket{1}\bra{1}$ are two orthogonal projectors on the
Hilbert space of the coin spanned by
$\{\ket{0}=(1,0)^\text{T},\ket{1}=(0,1)^\text{T}\}$,
$S\ket{x}=\ket{x+1}$ and $S^\dagger\ket{x}=\ket{x-1}$ are applied on
the walker's position. One can identify the eigenvectors $\ket{k}$
of $S$ and $S^\dagger$,
\begin{equation}
\ket{k}=\sum_x e^{ikx}\ket{x},
\end{equation}
with eigenvalues
\begin{equation}
S\ket{k}=e^{-ik}\ket{k}, S^\dagger\ket{k}=e^{ik}\ket{k}.
\end{equation}

Here a discrete-time QW is considered as a stroboscopic realization
of static effective Hamiltonian, defined via the single-step
evolution operator
\begin{equation}
U(\theta,\phi)=e^{-i H_\text{eff}(\theta,\phi)\delta t},
\end{equation}
where $\delta t$ is the time it takes to carry out one step and we
set $\delta t=1$ in the followings. The evolution operator for $N$
steps is given by $U^N(\theta,\phi)=e^{-i N
H_\text{eff}(\theta,\phi)}$. For the general rotations in
Eq.~(\ref{eq:rotation}), the effective Hamiltonian can be written as
\begin{equation}
H_\text{eff}(\theta,\phi)=\int_{-\pi}^{\pi}
\frac{\text{d}k}{2\pi}\left[\omega(k){\bf
n}(k)\cdot\bm{\sigma}\right] \otimes\ket{k} \bra{k},
\end{equation}
where the quasi-energy
\begin{align}
\omega(k)=&\pm \arccos\Big
\{\cos k\left[\cos\phi\cos\theta-(n_{x1}n_{x2}+n_{y1}n_{y2}+n_{z1}n_{z2})\sin\phi\sin\theta\right]\nonumber\\
& +\sin
k\left[n_{z2}\cos\phi\sin\theta+\sin\phi(n_{z1}\cos\theta+n_{y1}n_{x2}\sin\theta-n_{x1}n_{y2}\sin\theta)\right]\Big\},
\end{align}
and the unit vector ${\bm n}(k)=\left[n_x(k),n_y(k),n_z(k)\right]$
\begin{align}
&n_x(k)=\frac{1}{\sin\omega(k)} \Big\{-\sin k \left[n_{y2} \cos\phi
\sin\theta + \sin\phi(n_{y1} \cos\theta -
                  n_{z1} n_{x2} \sin\theta + n_{x1} n_{z2} \sin\theta)
                  \right]\nonumber\\
                  &+
      \cos k \left[n_{x2} \cos\phi \sin\theta +\sin\phi (n_{x1} \cos\theta
      +
                  n_{z1} n_{y2}\sin\theta - n_{y1} n_{z2}
                  \sin\theta)\right]\Big\},\nonumber\\
&n_y(k)=\frac{1}{\sin\omega(k)} \Big\{ (n_{y2} \cos k +
        n_{x2}\sin k)\cos\phi\sin\theta   + \cos
            k \sin\phi \left[n_{y1} \cos\theta + (-n_{z1} n_{x2} + n_{x1} n_{z2}) \sin\theta\right] \nonumber\\
            &+
        \sin k \sin\phi\left[n_{x1} \cos\theta + (n_{z1} n_{y2} -
                    n_{y1} n_{z2})
                    \sin\theta\right]\Big\},\nonumber\\
&n_z(k)=\frac{1}{\sin\omega(k)}\Big\{ (- \sin k \cos\phi+ n_{z1}\cos
k \sin\phi) \cos\theta +
 \sin k \sin\phi\sin\theta(n_{x1} n_{x2} + n_{y1} n_{y2} + n_{z1} n_{z2})\nonumber\\ & +
        \cos k (n_{z2}\cos\phi +n_{y1} n_{x2} \sin\phi - n_{x1} n_{y2}
        \sin\phi)\sin\theta
\Big\}.
\end{align}

The inverse Fourier transformation is
$\ket{x}=\int_{-\pi}^{\pi}\frac{\text{d}k}{2\pi} e^{-ikx}\ket{k}$.
The initial state of the walker+coin system can be written as
$\ket{\psi_0}=\ket{0}\otimes\ket{\Phi_0}$, where the original
position state of the walker is
$\ket{0}=\int_{-\pi}^{\pi}\frac{\text{d}k}{2\pi}\ket{k}$. In the $k$
basis, the evolution operator $U(\theta,\phi)$ becomes
\begin{equation}
U(\theta,\phi)\ket{k}\otimes\ket{\Phi_0}=\ket{k}\otimes
U_k(\theta,\phi)\ket{\Phi_0},
\end{equation}
where
\begin{equation}
U_k(\theta,\phi)=(e^{-ik}\mathcal{P}_0+e^{ik}\mathcal{P}_1)R_{\bf{n}_1}(\phi)R_{\bf{n}_2}(\theta)
=\left(
\begin{array}{cc}
A_{11} & A_{12} \\
A_{21} & A_{22} \\
\end{array}
 \right)
 \label{eq:Uk}
\end{equation}
is a $2\times2$ unitary matrix with the matrix elements
\begin{align}
&A_{11}=e^{-ik}\left[-(n_{x1}-in_{y1})(n_{x2}+in_{y2})\sin\phi\sin\theta+(\cos\phi+in_{z1}\sin\phi)(\cos\theta+in_{z2}\sin\theta)\right],\nonumber\\
&A_{12}=e^{-ik}\left[(in_{x1}+n_{y1})\sin\phi(\cos\theta-in_{z2}\sin\theta)+(in_{x2}+n_{y2})(\cos\phi+in_{z1}\sin\phi)\sin\theta\right], \nonumber\\
&A_{21}=e^{ik}\left[(in_{x1}-n_{y1})\sin\phi(\cos\theta+in_{z2}\sin\theta)+(n_{x2}+in_{y2})(i\cos\phi+n_{z1}\sin\phi)\sin\theta\right],\nonumber\\
&A_{22}=e^{ik}\left[-(n_{x1}+in_{y1})(n_{x2}-in_{y2})\sin\phi\sin\theta+(\cos\phi-in_{z1}\sin\phi)(\cos\theta-in_{z2}\sin\theta)\right].
\end{align}

At time $t$ (the time $t$ is proportional to the step number $N$),
the walker+coin state evolves to
\begin{equation}
\ket{\psi_t}=U^t(\theta,\phi)\ket{\psi_0}=\int_{-\pi}^{\pi}\frac{\text{d}k}{2\pi}\ket{k}\otimes
U_k^t(\theta,\phi)\ket{\Phi_0}.
\end{equation}
The probability for the walker to reach a position $x$ at time $t$
is
\begin{align}
p(x,t)&=\int_{-\pi}^{\pi}\frac{\text{d}k}{2\pi}\int_{-\pi}^{\pi}\frac{\text{d}k'}{2\pi}e^{-ix(k-k')}
\text{Tr}\left\{\left[U_{k'}(\theta,\phi)\right]^t\rho_0\left[U^\dagger_{k}(\theta,\phi)\right]^t\right\}\nonumber\\
&=\frac{1}{2}\int_{-\pi}^{\pi}\frac{\text{d}k}{2\pi}\left[\left(1+\bra{\Phi_0}{\bf
n}(k)\cdot{\bm\sigma}\ket{\Phi_0}\right)\delta(v_k-\frac{x}{t})+\left(1-\bra{\Phi_0}{\bf
n}(k)\cdot{\bm\sigma}\ket{\Phi_0}\right)\delta(v_k+\frac{x}{t})\right],
\end{align}
where $\rho_0=\ket{\Phi_0}\bra{\Phi_0}$, the group velocity of the
walker $v_k=\partial\omega(k)/\partial k$. To determine if there is
localization effect, we care more about the position variance and
the dependence of the variance on time. Thus we restrict our
interest to the moments of the distribution.
\begin{align}
\langle x^m \rangle_t&=\sum_x x^m p(x,t) =\sum_x
x^m\int_{-\pi}^{\pi}\frac{\text{d}k}{2\pi}\int_{-\pi}^{\pi}\frac{\text{d}k'}{2\pi}e^{-ix(k-k')}
\text{Tr}\left\{\left[U_{k'}(\theta,\phi)\right]^t\rho_0\left[U^\dagger_k(\theta,\phi)\right]^t\right\}.
\end{align}
With the formula of the delta function
$\sum_{x}x^me^{-ix(k-k')}/2\pi=i^m\delta^m(k-k')$, the expression of
the $m$th moment is rewritten as
\begin{align}
\langle
x^m\rangle_t&=\frac{i^m}{2\pi}\int_{-\pi}^{\pi}\text{d}k\int_{-\pi}^{\pi}\text{d}k'\delta^m(k-k')
\left[U_{k'}(\theta,\phi)\right]^t\rho_0\left[U^\dagger_{k}(\theta,\phi)\right]^t,\nonumber\\
&=i^m\int_{-\pi}^{\pi}\frac{\text{d}k}{2\pi}\frac{\text{d}^m\left[U_k(\theta,\phi)\right]^t}{\text{d}k^m}
\rho_0\left[U^\dagger_{k}(\theta,\phi)\right]^t.
\end{align}

Similar to a Hadamard coined walk~\cite{NV00}, one can find the
eigenvectors $\ket{\Phi_{kj}}$ of $U_k(\theta,\phi)$ and
corresponding eigenvalues $e^{\pm i\omega(k)}$. We can expand the
initial coin state $\ket{\Phi_0}=\sum_{j=1,2}c_{kj}\ket{\Phi_{kj}}$.
With $\frac{\text{d}}{\text{d}k}U_k(\theta,\phi)=-i\sigma_z
U_k(\theta,\phi)$~\cite{BCA03}, we only keep the diagonal
non-oscillatory terms and obtain
\begin{align}
\langle
x\rangle_t&=-\sum_{l=1}^{t}\int_{-\pi}^{\pi}\frac{\text{d}k}{2\pi}\sigma_z\left[U_k(\theta,\phi)\right]^l\rho_0\left[U_k^\dagger(\theta,\phi)\right]^l\nonumber\\
&=-t\int_{-\pi}^{\pi}\frac{\text{d}k}{2\pi}\sum_{j=1,2}|c_{kj}|^2\langle
\Phi_{kj}|\sigma_z|\Phi_{kj}\rangle+\text{oscillatory
terms}.\label{eq:x}
\end{align}
For non-degenerate unitary matrix $U_k(\theta,\phi)$, except for the
diagonal non-oscillatory terms, most of the terms are oscillatory,
which average to zero in the long-time limit~\cite{BCA03}.

Similarly, the second moment is obtained
\begin{align}
\langle
x^2\rangle_t=t^2\int_{-\pi}^{\pi}\frac{\text{d}k}{2\pi}\sum_{j=1,2}|c_{kj}|^2\langle\Phi_{kj}|\sigma_z|\Phi_{kj}\rangle^2
+\text{oscillatory terms}. \label{eq:sq}
\end{align}

From Eqs.~(8) and (\ref{eq:Uk}), we can see the spectrum $e^{\pm
i\omega(k)}$ of $U_k(\theta,\phi)$ is non-degenerate. Even for
degenerate $U_k(\theta,\phi)$ one can modify Eqs.~(\ref{eq:x}) and
(\ref{eq:sq}) to include appropriate cross terms, which does not
change the dependence of the position variance on time.

Generically, in the long-time limit, for a unitary coin the first
moment of the QW undergoes a linear drift and the variance grows
quadratically with time. There is a special case---the $\sigma_x$
coined QW, i.e., $R_{{\bf n}_1}(\phi)R_{{\bf
n}_2}(\theta)=\sigma_x$, in which the eigenstates of $U_k$ are
$\ket{\Phi_{k1}}=(-e^{-ik}\ket{0}+\ket{1})/\sqrt{2}$ and
$\ket{\Phi_{k2}}=(e^{ik}\ket{0}+\ket{1})/\sqrt{2}$, resulting in
$\langle\Phi_{kj}|\sigma_z|\Phi_{kj}\rangle=0$ (for $j=1,2$). Thus
the variance of the $\sigma_x$ coined QW does not depend on time.

In the two rotations case, the combination operation of two
rotations on the coin $R_{{\bf n}_1}(\phi)R_{{\bf n}_2}(\theta)$
shown in Eq.~(\ref{eq:Uk}) is unitary. Thus for arbitrary choices of
parameters $\theta$ and $\phi$ the position variance of the QW with
two time-independent rotations on the coin grows quadratically and
the behaviour of the QW is ballistic. Therefore, a second coin
rotation does not change the behaviour of QW from a ballistic spread
to localization.

The asymptotic analysis of the behaviour of this QW with two
time-independent coin rotations can be extended to more general QW
with more time-independent rotations on the coin. Once the combined
operator of the multiple-rotation on the coin is unitary, the
position variance grows quadratically with time and this QW shows
ballistic behaviour. No localization effect occurs.

This walk is homogeneous in either spatial or temporal space. The
coin rotations do not cause inhomogeneity in this walk which usually
leads to interesting localization effect. Furthermore, there is
another interesting idea~\cite{PG+14} on observation of topological
structures in this QW with multiple-rotation on the coin. The
signature of non-trivial topological structure is localization
effect due to localized bound states hosted at the topological
boundaries. However, based on our result on this walk, we can prove
that no localization occurs in this walk. That is because neither
topological boundary nor phase transition line exists in this walk.
The quasi-energy gap only closes at a few points in the parameter
space $(\theta,\phi)$. Here without losing generality, we use the
case of the QW in~\cite{PG+14} as an example. The unitary operator
of single-step of the QW with two consecutive non-commuting
rotations along $y$ and $x$ axes becomes
$U(\theta,\phi)=TR_x(\phi)R_y(\theta)$. The quasi-energy $\omega(k)$
becomes
\begin{equation}
\omega(k)=\pm\arccos\left[\cos
k\cos\theta\cos\phi-\sin k\sin\theta\sin\phi\right]
\end{equation}
and the unit vector is
\begin{equation}
{\bf n}(k)=\frac{1}{\sin\omega(k)}\left(
                                    \begin{array}{c}
                                      \sin\phi\cos\theta\cos k-\cos\theta\sin\phi\sin k \\
                                      \cos\phi\sin\theta\cos k+\sin\phi\cos\theta\sin k \\
                                      \sin\phi\sin\theta\cos k-\cos\phi\cos\theta\sin k \\
                                    \end{array}
                                  \right).
\end{equation}
The operator $U_k(\theta,\phi)$ is non-degenerate and unitary. Thus
the behaviour of this walk is ballistic as we show above. From the
dispersion relation we can see that this QW for generic $\theta$ and
$\phi$ has gaps around $\omega=0$ and $\omega=\pm\pi$. Therefore the
gap round $\omega=0$ closes at $k=0$, $(\theta=0,\phi=0)$ and
$(\theta=\pm\pi,\phi=\pm\pi)$, $k=\pi$, $(\theta=0,\phi=\pm\pi)$ and
$(\theta=\pm\pi,\phi=0)$, $k=\pi/2$,
$(\theta=\pi/2(-\pi/2),\phi=-\pi/2(\pi/2))$, and $k=-\pi/2$,
$(\theta=\pi/2(-\pi/2),\phi=\pi/2(-\pi/2))$, and the gap around
$\omega=\pm\pi$ closes at $k=0$, $(\theta=0,\phi=\pm\pi)$ and
$(\theta=\pm\pi,\phi=0)$, $k=\pi$, $(\theta=0,\phi=0)$ and
$(\theta=\pm\pi,\phi=\pm\pi)$, $k=\pi/2$,
$(\theta=\pi/2(-\pi/2),\phi=\pi/2(-\pi/2))$, and $k=-\pi/2$,
$(\theta=\pi/2(-\pi/2),\phi=-\pi/2(\pi/2))$. The gap closes at $13$
individual points in the parameter space $(\theta,\phi)$. There is
neither topological boundary nor phase transition line in the
parameter space. Thus no localized bound state hosted at the
topological boundaries exists in this walk.

In summary, we study the QW with two time-independent rotations on
the coin through the analytical solutions for the time dependence of
the position variance. The asymptotic result can be extended to the
walk with multiple time-independent rotations on the coin. As long
as the combination of the multi-rotations is unitary, the variance
grows quadratically with time and the QW shows ballistic behaviour.
No localization effect is observed in this QW. Although the fact
that two topics---QWs and localization effect meet, is fascinating
and opens the door to rich theoretical and experimental
investigation of quantum phenomena. Thus not only the investigation
on simulating localization with QWs but also the study on the
limitations on localization in quantum walk are important and worthy
of attention. Our research exactly gives insight into limitations on
localization.

\acknowledgements The authors thank Yongsheng Zhang and Shunlong Luo
for simulating discussions. This work has been supported by NSFC
under 11174052, 973 Program under 2011CB921203 and the Open Fund
from the SKLPS of ECNU.

\end{document}